\documentclass[twocolumn]{aastex631}

\usepackage{amsmath}	
\usepackage{amssymb}
\usepackage{tabularx}

\newcommand{\photoz}{photo-$z$}
\newcommand\eazy{\textsc{Eazy}}
\newcommand{\Lephare}{\textsc{LePhare}}
\newcommand{\palpha}{{\textsc{Prospector}-$\alpha$}}

\newcommand{\bl}[1]{{#1}}

\mathchardef\mhyphen="2D
\newcommand{\secref}[1]{{}Sec.~\ref{#1}}

\newcommand{\eqn}[1]{\begin{align}#1\end{align}}
\newcommand{\ie}{{\textit{i.e.},~}}
\newcommand{\eg}{{\textit{e.g.},~}}
\newcommand{\equref}[1]{{}Eq.~(\ref{#1})}
\newcommand{\figref}[1]{{}Fig.~\ref{#1}}

\submitjournal{APJS}

\begin{document}

\title{Hierarchical Bayesian inference of photometric redshifts with stellar population synthesis models}

\correspondingauthor{Boris Leistedt}
\email{b.leistedt@imperial.ac.uk}

\author[0000-0002-3962-9274]{Boris Leistedt}
\affiliation{Department of Physics, Imperial College London, Blackett Laboratory, Prince Consort
Road, London SW7 2AZ, UK}

\author[0000-0003-4618-3546]{Justin Alsing}
\affiliation{Oskar Klein Centre for Cosmoparticle Physics, Department of Physics, Stockholm University, Stockholm SE-106 91, Sweden}

\author[0000-0002-2519-584X]{Hiranya Peiris}
\affiliation{Department of Physics and Astronomy, University College London, Gower Street, London, WC1E 6BT, UK}
\affiliation{Oskar Klein Centre for Cosmoparticle Physics, Department of Physics, Stockholm University, Stockholm SE-106 91, Sweden}

\author[0000-0002-0041-3783]{Daniel Mortlock}
\affiliation{Department of Physics, Imperial College London, Blackett Laboratory, Prince Consort
Road, London SW7 2AZ, UK}
\affiliation{Department of Mathematics, Imperial College London, London SW7 2AZ, UK}
\affiliation{Oskar Klein Centre for Cosmoparticle Physics, Department of Physics, Stockholm University, Stockholm SE-106 91, Sweden}

\author[0000-0001-6755-1315]{Joel Leja}
\affil{Department of Astronomy \& Astrophysics, The Pennsylvania State University, University Park, PA 16802, USA}
\affil{Institute for Computational \& Data Sciences, The Pennsylvania State University, University Park, PA, USA}
\affil{Institute for Gravitation and the Cosmos, The Pennsylvania State University, University Park, PA 16802, USA}

\begin{abstract}
We present a Bayesian hierarchical framework to analyze photometric galaxy survey data with stellar population synthesis (SPS) models.
Our method couples robust modeling of spectral energy distributions with a population model and a noise model to characterize the statistical properties of the galaxy populations and real observations, respectively.
By self-consistently inferring all model parameters, from high-level hyper-parameters to SPS parameters of individual galaxies, one can separate sources of bias and uncertainty in the data.
We demonstrate the strengths and flexibility of this approach by deriving accurate photometric redshifts for a sample of spectroscopically-confirmed galaxies in the COSMOS field, \bl{all with 26-band photometry and spectroscopic redshifts. We achieve} a performance competitive with publicly-released photometric redshift catalogs based on the same data.
Prior to this work, this approach was computationally intractable in practice due to the heavy computational load of SPS model calls; we overcome this challenge using with neural emulators.
We find that the largest photometric residuals are associated with poor calibration for emission line luminosities and thus build a framework to mitigate these effects.
This combination of physics-based modeling accelerated with machine learning paves the path towards meeting the stringent requirements on the accuracy of photometric redshift estimation imposed by upcoming cosmological surveys. 
The approach also has the potential to create new links between cosmology and galaxy evolution through the analysis of photometric datasets.
\end{abstract}

\keywords{photometric redshifts - galaxy surveys - cosmological parameters}

\section{Introduction} \label{sec:intro}

Deriving accurate redshifts and redshift distributions from photometry alone is of central importance for the scientific exploitation of modern galaxy surveys.
In particular, cosmological analyses involving galaxy clustering and weak gravitational lensing require exquisitely accurate estimates of the redshift distributions of the selected galaxy populations (see \eg \citealt{Hildebrandt_2020, Abbott_2022}).
Photometric redshifts ({\photoz}'s) of individual galaxies are also useful for selecting targets of interest (\eg at high redshift).

Photo-$z$'s are typically obtained via two types of methods: template fitting, and machine learning\footnote{A third technique, clustering redshifts, can constrain redshift distributions, but cannot deliver individual {\photoz}'s, see \eg \cite{Schneider_2006, Newman_2008, menard_2013, McQuinn_2013, Schmidt_2013, Morrison_2017, Gatti_2021}.}.
Template fitting (\eg, \citealt{Benitez_2000, Ilbert_2006, Brammer_2008, Ilbert_2006, Tanaka_2015, Acquaviva_2015, Carnall_2018, Battisti_2019}) relies on lists of grids or templates for the spectral energy distributions (SEDs) of the galaxies considered, combined with a prior on the parameters involved (galaxy type, magnitude, etc). 
Machine learning techniques (\eg \citealt{Collister_2004, Carrasco_Kind_2013}) involve flexible models trained on spectroscopic or synthetic data.
Despite their past successes, both these approaches are unable to satisfy the accuracy requirements imposed by upcoming Stage IV surveys, as they are not robust to errors in the data (calibration offsets, underestimated uncertainties), in the SED model, or to the non-representativity or sample variance of the spectroscopic data used for training and validation (\eg \citealt{Hartley_2020, Gatti_2021, Myles_2021, Newman_2022}).
For example, the Vera C. Rubin Observatory's  Legacy Survey of Space and Time (LSST)  will require the uncertainty on the mean of redshift distributions to be smaller than $0.001(1+z)$ by its Year 10 data release \citep{lsst_sciencereq}, which in turn imposes very stringent requirements on the accuracy of individual {\photoz}'s.

We introduce a framework capable of addressing these needs: a hierarchical model of galaxy photometry, consisting of three components describing a generating process of survey data: 
\begin{itemize}
    \item galaxy SEDs generated with stellar population synthesis;
    \item a population model describing the distributions of the intrinsic parameters of galaxies in the sample analyzed;
    \item a noise model for generating photometric observations (\ie a likelihood function).
\end{itemize}
This provides a powerful combination of empirical modeling with physics-driven components, incorporating known (astro)physics while leaving sufficient room for data-driven corrections, addressing the aforementioned issues of flexibility, interpretability, mis-calibration, and extrapolation outside of training data. 
Thus, once applied and calibrated to large survey data-sets, it has the potential to yield both precise and accurate inferences on intrinsic galaxy parameters, including redshift.
Building on previous work on hierarchical modeling in photometric surveys \citep{Leistedt_2016, Jones_2018, S_nchez_2018, Rau_2019, Alarcon_2020, S_nchez_2020}, our framework takes advantage of the increasing accuracy and flexibility of SPS, and of methodological innovations in Bayesian inference for large hierarchical models, in particular involving machine learning-facilitated acceleration of computationally-intensive analysis steps.

In this paper, we demonstrate the efficacy of this framework by applying it to the COSMOS2020 data set \citep{Weaver_2022}.
For the choice of SED model, we consider \palpha~\citep{Leja_2017, Leja_2019, Johnson_2021}.
We address the high computational needs of its model calls with neural emulators \citep{Alsing_2020} applied to both SPS total fluxes and emission line contributions. 
We find that the emission line contributions are responsible for the largest photometric residuals in our analysis and therefore require calibration in order to derive accurate \photoz's.
Together, the SED, data, and population models form a complex hierarchical model which would normally be computationally intractable, a challenge we overcome with neural emulator-based acceleration.
We obtain redshift estimates competitive with those publicly released with the COSMOS2020 data using this baseline model, providing a benchmark validation test demonstrating the power of the new methodology.
The extensive \bl{26-band} wavelength coverage of COSMOS2020 simplifies parameter inference and allows us to demonstrate the conceptual advantages of the new methodology without the significant overhead of carrying out a Markov chain Monte Carlo (MCMC) analysis.
\bl{Application to broadband-only photometric surveys, such as the LSST, will require a more involved treatment of the uncertainties and selection effects, which we defer to future work.}
In a companion paper \citep{Alsing_2022}, we show how to correctly include selection effects in the forward modeling of \bl{such} surveys, and in particular for the inference of galaxy redshift distributions.

The outline of this paper is as follows. In  \secref{sec:generative} we describe our framework in the form of a generative model of photometric data. 
An inference formalism is presented in \secref{sec:inference}.
We describe the data in \secref{sec:data}, and the results in \secref{sec:results}. 
We discuss the results in further detail in \secref{sec:discussion}, and conclude in \secref{sec:conclusion}.
In what follows, a Planck 2015  \citep{planck2015} and a \cite{Chabrier_2003} initial mass function are adopted for all relevant calculations.
$p(\cdot|\cdot)$  refers  to  the  probability density of the quantity before the “$|$”, conditioned on quantities behind it.
We work with fluxes in AB units.

\section{Generative model for galaxy photometry} \label{sec:generative}

Hierarchical models tackling photometric redshifts and redshift distributions have been developed \citep{Leistedt_2016, Rau_2019, Jones_2018, S_nchez_2018, Alarcon_2020, S_nchez_2020}, but typically assume a fixed SED model.
\cite{Leistedt_2019} pioneered the approach of inferring redshifts jointly with calibration of the hyper-parameters of a SED model.
However, like most \photoz~methodologies \citep{Brammer_2008, Ilbert_2006, Tanaka_2015, Acquaviva_2015, Carnall_2018, Battisti_2019}, the model still relied on a curated set of SED templates, which made it difficult to construct more structured, interpretable corrections to the SEDs or to the population model.
Our work introduces the use of full continuous stellar population synthesis (SPS) models for redshift estimation, tailored to the multi-level calibration and inference that will be needed for the upcoming generation of wide-deep photometric surveys \citep{Newman_2022}.

\begin{table*}
\centering
\scalebox{0.95}{
\begin{tabularx}{\textwidth}{cc}
\toprule
Parameter & Description \tabularnewline
\hline \tabularnewline
& \emph{hyper-parameters} \tabularnewline

$\omega_b$ & Zero-point offset (for $b$'th band, relative to the $i$ band) 
\tabularnewline
$\gamma_b$ & Additional flux uncertainty contribution for the $b$'th band (fraction of the total model flux)
\tabularnewline
$\alpha_j$ & Additional flux for the $j$'th emission line (fraction of the line strength in \textsc{FSPS})
\tabularnewline
$\beta_j$ & Additional flux uncertainty contribution for the $j$'th emission line (fraction of the line strength in \textsc{FSPS})
\tabularnewline 
$\boldsymbol{\kappa}$ & hyper-parameters describing the galaxy population model \tabularnewline\tabularnewline

& \emph{Latent parameters} \tabularnewline

$\boldsymbol{\varphi}$ & Stellar population parameters describing the rest-frame spectrum (per galaxy) \tabularnewline

$z$ & Redshift (per galaxy) \tabularnewline \tabularnewline

& \emph{Derived parameters} \tabularnewline

$\ell_\mathrm{SPS}(\boldsymbol{\varphi})$ & Flux energy density predicted by \textsc{FSPS} (per galaxy)
\tabularnewline 
$\ell_j(\boldsymbol{\varphi})$ & Amplitude of emission line (Dirac delta model) in \textsc{FSPS} (per galaxy)
\tabularnewline 
$L_{jb}(\boldsymbol{\varphi}, z)$ & Flux of the $j$'th emission line in the $b$'th band (per galaxy), \ie 
$\ell_j$ inserted in \equref{eq:fluxdefinition} \tabularnewline 
$F_b(\boldsymbol{\varphi}, z, \boldsymbol{\alpha})$ & Total model flux in the $b$'th band, with all contributions (per galaxy), \ie 
\equref{eq:modelfluxes} inserted in \equref{eq:fluxdefinition}
\tabularnewline
$\Sigma_b(\boldsymbol{\varphi}, z, \boldsymbol{\alpha}, \gamma_b, \boldsymbol{\gamma})$ &  Additional flux uncertainty in the $b$'th band, with all contributions (per galaxy), see
\equref{eq:extrauncertainty}
\tabularnewline\tabularnewline

& \emph{Data} \tabularnewline

$\hat{F}_b$ & Measured flux in the $b$'th band (per galaxy)\tabularnewline

$\sigma_b$ & Flux measurement uncertainty in the $b$th band (per galaxy)
\tabularnewline

\tabularnewline

\hline
\end{tabularx}
}
\caption{Notation for all model parameters. Vectors are indicated with bold symbols. Our total log likelihood and graphical model presented below include an extra index $i$ per galaxy, which we have omitted in this table and all other equations in the paper for simplicity.}
\label{tab:parameters}
\end{table*}

\subsection{SED model}\label{sec:sedspsmodel}

SPS provides a powerful way to model the SED of a galaxy.
It exploits fundamental principles to generate and sum the contributions of `simple' stellar populations (as observed in, \eg star clusters) and apply the effects of additional complications such as dust and nebular emission in order to create galaxy SEDs within a forward modeling framework.
This step involves a set of specific modeling choices for these components, for example an explicit star formation history (SFH).
We use the Flexible Stellar Population Synthesis (\textsc{FSPS}, \citealt{Conroy_2009, Conroy_2009b, Conroy_2010}) code, accessed through the \textsc{python-FSPS} binding \citep{dan_foreman_mackey_2014_12157}.

For the SED model we consider a variation of \palpha, which was first introduced in \cite{Leja_2017}, revised in \cite{Leja_2019, Johnson_2021}, and has been successfully used to constrain Bayesian models for the galaxy population \cite{Leja_2020, leja_2022, Nagaraj_2022, Whitler_2022}.
While these works fixed the redshift to the value measured with spectroscopy (spec-$z$), we relax this assumption and treat it as a free parameter.
In our variation of \palpha, a galaxy SED is (deterministically) defined by 15 parameters:
\begin{itemize}
    \item 8 parameters describing the SFH: 1 parameter for the total stellar mass formed $\log M_\star$  [M$_{\odot}$] (integral of the SFH); 1 parameter for the stellar metallicity $\log(Z_\star/Z_\odot)$, assumed to be the same for all stars in the galaxy; and 6 parameters for the relative age SFH bins (ratios $r_i$ of the piece-wise SFR in adjacent temporal bins), where the seven time bins are spaced following \cite{Leja_2020}.
    \item 3 parameters for the dust attenuation, following the model of \cite{Charlot_2000} (see \citealt{Leja_2017} for details) with birth-cloud $\tau_1$, and diffuse attenuation $\tau_2$ with a power law (of index $n$) from \cite{calzetti2000}.
    \item a gas-phase metallicity parameter $\log(Z_\mathrm{gas}/Z_\odot)$ for the nebular emission (decoupled from the stellar metallicity) modeled after the grids of \cite{Byler_2017}. 
    \item two parameters, $\log{\rm f}_{\rm AGN}$ and $\tau_{\rm AGN}$, for the AGN torus emission model of \cite{Nenkova_2008}.
    \item redshift, $z$
\end{itemize}
The SED includes dust heating from stars via energy balance, via a dust SED of fixed shape \citep{Draine_2007}.

\subsection{Emission lines}\label{sec:emlines}

\palpha~includes a nebular emission model where the gas is ionized by the same stars synthesized in the SED  \citep{Byler_2017, nell_byler_2018_1156412}.
There is a large variability of the line strengths, as demonstrated by \cite{Byler_2017}.
In practice, previous studies have required the flexible addition of  emission lines on top of SED templates in order to obtain accurate \photoz's \citep{Ilbert_2006, Ilbert_2008, Brammer_2008, Alarcon_2021}, so we adopt a similar approach. 

We model offsets and uncertainty in the strength of the emission lines by encoding them into parametric bias and variance parameters.
For a set of SPS parameters $\boldsymbol{\varphi}$, the total energy density as a function of wavelength, $\ell$, consists of the base \textsc{FSPS} prediction $\ell_\mathrm{SPS}$, with an extra additive contribution from emission lines.
In practice, in \textsc{FSPS}, emission lines are modeled as a delta functions, integrated in band-passes, and added to the model photometry.
We can therefore write the SED as
\eqn{
    \ell(\lambda; \boldsymbol{\varphi}, \boldsymbol{\alpha})\ =\ \ell_\mathrm{SPS}(\lambda; \boldsymbol{\varphi}) 
    + \sum_{j=1}^{\mathrm{N}_\mathrm{lines}} \alpha_j \ell_j(\boldsymbol{\varphi}) 
    \delta^\mathrm{D}(\lambda-\lambda_j)\quad \
\label{eq:modelfluxes}
}
where $\ell_j$ is the amplitude of the $j$'th line, $\lambda_j$ the rest-frame wavelength of the line, and $\delta^\mathrm{D}(\cdot)$ is the Dirac delta function.
$\alpha_j$ the hyper parameter for the additional contribution from line $j$. 
Because line $\ell_j$ is already included in $\ell_\mathrm{SPS}$ with a default weight, its  total contribution in $\ell$ is $1+\alpha_j$.

\subsection{Synthetic photometry}\label{sec:sedmodel}

We can obtain total model fluxes in the $b$'th band by integrating the SED through the band-pass filter $W_b(\lambda)$, such that
\eqn{
    & F_b(\boldsymbol{\varphi}, z, \boldsymbol{\alpha}) = \\
    & \quad\quad \frac{(1+z)^{-1}}{4\pi d_L^2(z)}\int_0^\infty \ell\Bigl(\frac{\lambda}{1+z}; \boldsymbol{\varphi}, \boldsymbol{\alpha}\Bigr)\,
    e^{-{\tau}(z, \lambda)}
    W_b(\lambda)d\lambda,\nonumber \quad \quad
    \label{eq:fluxdefinition}
}
where 
$d_L(z)$ the luminosity distance, and ${\tau}(z, \lambda)$ the effective optical depth of the inter-galactic medium, calculated with the model of \cite{1995ApJ...441...18M}.

\subsection{Population model}

So far we have a mechanism for using SPS to generate the photometry of a single galaxy given its intrinsic properties, $\boldsymbol{\varphi}$, and redshift, $z$.
We now describe a formalism for generating a sample of galaxies with a probability distribution $p(\boldsymbol{\varphi}, z | \boldsymbol{\kappa})$, where  $\boldsymbol{\kappa}$ are the hyper-parameters which describe the population.
These could in principle be inferred as well, but this typically requires a careful treatment of selection effects (\ie how galaxies are selected into the sample at hand), which must be included at the population level. We show how to treat selection effects in a companion paper, \cite{Alsing_2022}.
For our demonstration with COSMOS2020 in this paper, we focus on redshift inference for individual galaxies. Due to the constraining power of the data, the population prior will have very little impact on the galaxy posterior distributions, especially the redshifts. 
Therefore, we are able to bypass an explicit treatment of selection effects, and also adopt a fairly uninformative population model.

We adopt minor variations on the fiducial \palpha~prior (see \citealt{Leja_2019, Johnson_2021}), summarized in Table~\ref{tab:spsparampriors}.
It factorizes as
\begin{align}
   & p(\boldsymbol{\varphi}, z | \boldsymbol{\kappa}) = p(\boldsymbol{\varphi} | z, \boldsymbol{\kappa}) \ p(z)    \\
    &  \quad\ \ =\  p(\mathrm{mass}) \times p(\mathrm{SFH} ) \times p(\mathrm{dust}| \boldsymbol{\kappa}) \nonumber \\
    &  \quad\quad \times p(\mathrm{stellar\ metallicity}) \times p(\mathrm{AGN}) \times p(z)  \nonumber \\
    &  \quad\quad  \times p(\mathrm{gas\mhyphen phase\ metallicity} |\mathrm{stellar\ metallicity, \boldsymbol{\kappa}}) \, . \nonumber
\end{align}
Most parameters have simple uniform or log-uniform priors, except in the cases discussed below.

\begin{deluxetable}{lll}[ht]
\tablewidth{0.43\textwidth}
\tablecaption{Summary of parameters and priors describing the SPS model. \label{tab:spsparampriors} $\mathcal{N}_c(l, s, m, M)$ refers to a truncated normal distribution of location $l$, scale $s$, and in the range $[m, M]$. Our fiducial values for the parameters below are $\mu_n = -0.095 + 0.111\, \tau_2- 0.0066\, \tau_2^2$, $\sigma_n=0.4$, $\sigma_Z=3.0$, $(\mu_1, \sigma_1) = (1.0, 0.3)$, $(\mu_2, \sigma_2) = (0.3, 1.0)$, following \cite{Leja_2019}.
}
\tablehead{
\colhead{Parameter}& \colhead{Prior bounds}  & \colhead{Prior}
}
\startdata
$\log M_\star$  [M$_{\odot}$]        & $[7, 13]$ & Uniform($7$, $13$)   \\
$\log(Z_\star/Z_\odot)$                             & $[-1.98, 0.19]$& Uniform($-1.98, 0.19$) \\
$\{ \log r_i \}_{i=1, \cdots, 6}$                           & $[-5, 5]$ & Stu($0, 0.3, 2$) \\
$\tau_2$                             & $[0, 4]$& $\mathcal{N}_c(\mu_2, \sigma_2, 0, 4)$ \\
$\tau_1/\tau_2$                                 & $[0, 2]$& $\mathcal{N}_c(\mu_1, \sigma_1, 0, 2)$ \\
$n$               & $[-1, 0.4]$& $\mathcal{N}_c(\mu_n, \sigma_n, -1, 0.4)$ \\
$\log{\rm f}_{\rm AGN}$                               & $[10^{-5}, 150]$ & LogUniform($10^{-5}, 150$) \\
$\tau_{\rm AGN}$                                      & $[-2, 0.5]$ & LogUniform($5, 150$) \\
$\log(Z_\mathrm{gas}/Z_\odot)$                             & $[-2, 0.5]$& $\mathcal{N}_c(\log(Z_\star/Z_\odot), \sigma_Z, -2, 0.5)$ \\
$z$                                      & $[0, 2.5]$ & Uniform($0, 2.5$) \\
\enddata
\end{deluxetable}

For the gas-phase metallicity we take a normal\footnote{Since we apply normal priors to parameters only defined in finite ranges, a renormalization is necessary, giving rise to truncated normals, as described in Table~\ref{tab:spsparampriors}. For simplicity we omit this detail throughout this section.} prior of mean at the $\mathrm{stellar\ metallicity}$ and standard deviation $\sigma_Z = 3.0$.
Despite the fact that they should track each other approximately, it is more robust to treat the gas-phase metallicity as a nuisance parameter \citep{Leja_2020}.
We take a normal prior on the diffuse dust components $\tau_2$ with mean equal to $\mu_2 = 0.3$ and standard deviation $\sigma_2 = 1.0$.
For the birth cloud component $\tau_1$ we take a normal prior on the ratio $r = \tau_1 / \tau_2$, with mean equal to $\mu_1 = 1$ and standard deviation $\sigma_1 = 0.3$.
The index of the dust attenuation law (for the diffuse component) is assumed to vary as a function of the total dust attenuation, with mean given by:
$\langle \delta \rangle = \mu_n = -0.095 + 0.111 \,\tau_2- 0.0066\, \tau_2^2,$
where $\delta$ is the (negative) offset from the index of the Calzetti attenuation curve \citep{calzetti2000}.
We take a normal prior on $\delta$, with mean $ \mu_n$ given above and standard deviation $\sigma_\delta = 0.4$.
This is a simple average of the results of \cite{Leja_2019}, but could be updated with a more complex prior, such as that of \cite{Nagaraj_2022}.

Finally, we collect the following hyper-parameters in a vector $\boldsymbol{\kappa} = (\mu_r, \sigma_r, \sigma_n, \sigma_\delta, \sigma_Z)$.
We will infer them jointly with the other components of the model, in order to demonstrate how one could in principle learn about the underlying physics with this type of forward modeling framework. 
However, we expect little sensitivity to these hyper-parameters for the dataset we use.
A more detailed population prior, encoding more of the known relations from galaxy formation and evolution, was developed by \cite{Alsing_2022}, and successfully reproduced galaxy redshift distributions given broad-band photometry.

\subsection{Noise model}

The final stage of the generative model is to simulate observed photometry $\hat{F}_b$ from the model flux $F_b$, by adding noise.
The noise model can be fully characterized using a likelihood function. We use a scaled and translated Student's-t distribution\footnote{The density of the scaled and translated Student's-t distribution is 
\eqn{
    \mathrm{Stu}(x; l, s, \nu) =  \frac{\Gamma(\frac{\nu+1}{2})}
    {\sqrt{\nu\pi} \Gamma(\frac{\nu}{2})} \left( 1 + \frac{t^2}{\nu}\right)^{-\frac{\nu+1}{2}}\nonumber
    }
with $t=(x-l)/s$ the residuals and $\Gamma(\cdot)$ the Gamma function. $l$ and $s$ are often referred to as location and scale parameters, and $\nu$ the number of degrees of freedom.} with two degrees of freedom, which is more robust to outliers than a normal distribution, since it has heavier tails:
\eqn{
    \label{eq:likelihood}&p(\hat{F}_b | F_b, \sigma_b, \Sigma_b, \omega_b) = \\ 
    & \quad\quad \mathrm{Stu}\bigl( 
    \hat{F}_b\ ;
    \ \omega_b F_b\ ,
    \ \sigma_b^2 + (\omega_b\Sigma_b)^2\ , 
    \ 2
    \bigr) .\nonumber 
}
Aside from the observational uncertainty $\sigma_b$, which is given for each object (typically measured from the images), we introduce two additional terms: a zero-point offset $\omega_b$, and the model uncertainty $\Sigma_b$ added in quadrature.
$\omega_b$ appears in front of the uncertainty because the latter will involve the model $F_b$, which must be rescaled by $\omega_b$ (although there is some freedom in choosing how to define these terms self-consistently).

We write the model uncertainty as the sum (in quadrature) of a fraction of the total model flux and a fraction of the total line fluxes,
\eqn{
    &\Sigma_b^2(\boldsymbol{\varphi}, z, \boldsymbol{\alpha}, \gamma_b, \boldsymbol{\gamma}) = \label{eq:extrauncertainty}  \\
    & \quad \left(\gamma_b  F_b(\boldsymbol{\varphi}, z) \right)^2+ \sum_{j=1}^{\mathrm{N}_\mathrm{lines}} \left(\beta_j (1 + \alpha_j) L_{jb}(\boldsymbol{\varphi}, z) \right)^2, \nonumber
}
where $L_{jb}$ is the photometric flux of the $j$'th line in the $b$'th band, \ie the result of applying \equref{eq:fluxdefinition} to $\ell_j \delta^\mathrm{D}(\lambda-\lambda_j)$.
The first term operates at the level of the band (controlled by a hyper-parameter $\gamma_b$) and is analogous to the magnitude uncertainty floor often employed in {{\photoz}} techniques.
The second term is similar but adds a contribution from each line. 
As before, the hyper-parameter $\beta_j$ can be interpreted in terms of a fraction of the total emission line flux (which is now $1+\alpha_j$). 
In summary, this allows us to separate the extra uncertainty $\Sigma_b$ preferred by the data into contributions from calibration uncertainty (via $\gamma_b$) and SED uncertainty (via $\beta_j$).

\begin{figure}[h!]
    \centering
    \includegraphics[width=0.8\columnwidth]{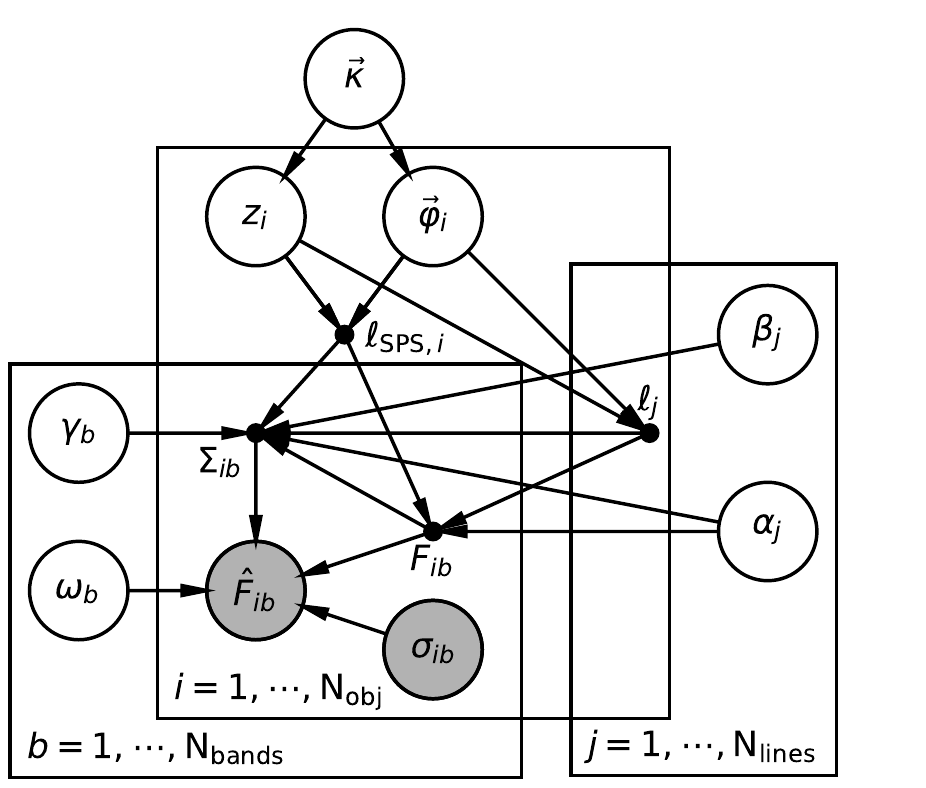}
    \caption{Graphical representation of the main parameters involved of our model. Circles are inferred random variables, shaded circles are observed, and dots indicate random variables that are deterministic in their inputs.}
    \label{fig:pgn}
\end{figure}

\section{Inference formalism}\label{sec:inference}

We now assemble the components presented in the previous section into a hierarchical model for generating and analysing photometric data.
We describe an inference methodology and also computational acceleration using machine learning.
\newpage
\subsection{Hierarchical model}

A summary of the parameters of our model is provided in Table~\ref{tab:parameters}. A graphical representation is provided in \figref{fig:pgn}.
Each galaxy has 15 intrinsic SPS parameters (including redshift), as described in \secref{sec:sedspsmodel}, and in Secs 2.2 to 2.5 we have introduced five sets of hyper-parameters, aiming to add some targeted sources of flexibility in the model.
Furthermore, a hierarchical modeling approach has the ability to fix subsets of parameters, and to observe the effect on the others, offering additional robustness diagnostics.

The full posterior distribution of our model (adding a subscript $i$ for galaxies) is
\begin{widetext} \begin{equation}
    p\bigl(\underbrace{\boldsymbol{\alpha}, \boldsymbol{\beta}, \boldsymbol{\omega}, \boldsymbol{\gamma}, \boldsymbol{\kappa}}_\mathrm{hyper}, \underbrace{\{\boldsymbol{\varphi}_i, z_i\}}_\mathrm{latent}\bigr) = 
    \underbrace{
    p(\boldsymbol{\alpha}, \boldsymbol{\beta}, \boldsymbol{\omega}, \boldsymbol{\gamma}, \boldsymbol{\kappa})
    }_\mathrm{global\ prior}
    \prod_{i=1}^{\mathrm{N}_\mathrm{obj}} \underbrace{p(\boldsymbol{\varphi}_i, z_i | \boldsymbol{\kappa})}_\mathrm{population\ model} \prod_{b=1}^{\mathrm{N}_\mathrm{bands}} \underbrace{p\Bigl(\hat{F}_{ib} \Bigl| \underbrace{ F_{ib}(\boldsymbol{\varphi}_i, z_i, \boldsymbol{\alpha})}_\mathrm{SED\ model}, \sigma_{ib}, \Sigma_{ib}^2(\boldsymbol{\varphi}_i, z, \boldsymbol{\alpha}, \gamma_b, \boldsymbol{\gamma}), \omega_b\Bigr)}_\mathrm{noise\ model} \, .\label{eq:fullposterior}
\end{equation} \end{widetext}

\subsection{Priors}

In our approach it is straightforward to encode prior knowledge or constraints on corrections such as zero-points or emission line contributions as 
the prior on hyper-parameters appears explicitly as
$p(\boldsymbol{\alpha}, \boldsymbol{\beta}, \boldsymbol{\omega}, \boldsymbol{\gamma}, \boldsymbol{\kappa})$. 
The priors we use are given in Table~\ref{tab:hyper_priors}.

\begin{deluxetable}{lll}[ht]
\tablewidth{0.43\textwidth}
\tablecaption{Priors on hyper-parameters. The numbers in parentheses indicate the location and scale of the distribution.}
    \label{tab:hyper_priors}
\tablehead{
\colhead{Hyper-parameter} & \colhead{Prior}
}
\startdata
        $\{\alpha_j\}_{j=1, \cdots, \mathrm{N}_\mathrm{lines}}$ &  Laplace(0, 10) \\
        $\{\beta_j\}_{j=1, \cdots, \mathrm{N}_\mathrm{lines}}$ &  Normal(0, 1) \\
        $\{\omega_b\}_{b=1, \cdots, \mathrm{N}_\mathrm{bands}}$ & Normal(1, 1) \\
        $\{\gamma_b\}_{b=1, \cdots, \mathrm{N}_\mathrm{bands}}$ & Normal(0, 1) \vspace*{1mm}\\
\enddata
\end{deluxetable}

We find that the Laplace prior for each $\alpha_j$ is preferable to a normal prior since it promotes sparsity of the coefficients, \ie the coefficients should be zero unless they significantly improve the model fits.
Without this regularization there is a risk that line contributions would mimic other narrow features in the SED, such as absorption lines, which should instead be associated with a separate set of stellar evolution physics.

\subsection{Optimization strategy}\label{sec:optimization}

Our inference strategy is to optimize the posterior distribution with respect to all parameters. 
This is because the mixture of broad bands (typically deeper) and noisier intermediate or narrow bands makes it possible to pin down SED features accurately and unequivocally.
As a result, it is possible to derive estimates of the hyper- and latent parameters through optimization, without the need for more involved inference strategies, such as MCMC.
We adopt the \textsc{Adam} optimizer with a learning rate of $10^{-3}$.

Challenges that need to be overcome at this stage involve the computational needs of SPS calls, and the fact that they are not differentiable analytically (although differentiable SPS models are now being developed, \citealt{https://doi.org/10.48550/arxiv.2205.04273, https://doi.org/10.48550/arxiv.2112.06830}).
We address these issues with neural emulators of the model predictions, which are both $\sim 10^4$ faster than native SPS calls and differentiable \citep{Alsing_2020}. We return to the details of our neural emulators in the next section.

In principle, one could determine all parameters in a single run. 
But for the purpose of this demonstration study we adopt a two-stage  approach (\eg \citealt{Laigle_2016, Weaver_2022}) which is commonly used and leverages the availability of spectroscopic redshifts.
We first determine the hyper-parameters with redshifts fixed at the values acquired via spectroscopy (optimizing all SPS parameters other than the redshift).
A second run is then performed by fixing the hyper-parameters at the inferred values, and only optimizing the latent SPS parameters (now including the redshift).
The resulting optimized redshifts are denoted $\hat{z}^\mathrm{MAP}$ (omitting the galaxy index).

Finally, we employ bijectors to map the original parameters to new variables which have zero-mean, unit-variance normal priors.
These are easier to work with than the original variables since they do not have boundaries, which facilitates optimization (and sampling).
The parameter bounds are shown explicitly in Table~\ref{tab:spsparampriors}.
 
\subsection{Emulators}

As previously mentioned, one drawback associated with SPS models is the computational cost of each model call, which is the result of the many intensive numerical steps involved in predicting flux densities and photometry from stellar isochrones, initial mass function, and the other ingredients of SPS. While this has been overcome with MCMC in the past and applied to medium-sized ($\sim 10^5$ objects) data sets (\eg \citealt{Leja_2019}), photometric galaxy surveys are approaching of order a billion objects \citep{Abbott_2022, lsst_sciencereq}, so brute-force inference is not practical. 
Hyper-parameters or model calibration would make the computational load even heavier. We now discuss how we employ emulators\footnote{Note that emulators are not the only approach to accelerating inference.
For example, \cite{anpe} and \cite{sythz} showed how to approximate posterior distributions directly for the estimation of SPS parameters or redshifts from broad-band photometry.} of fluxes and line emission in order to overcome this challenge.

\subsubsection{SPS fluxes}

For each band, we use the methodology described in \cite{Alsing_2020} to train an emulator to predict the photometric flux, \ie $\ell_\mathrm{SPS}$ fed into \equref{eq:fluxdefinition}.
While in principle one could train emulators to predict the $\ell_\mathrm{SPS}(\lambda)$, it is advantageous to emulate photometric fluxes directly in order to avoid evaluating \equref{eq:fluxdefinition}.
The emulator for each band is a fully-connected 4-layer neural network (with 128 neurons in each layer).
We train with staggered learning rates ($10^{-3}$, $10^{-4}$, $10^{-5}$, $10^{-6}$) and batch sizes (1000, 10000, 50000, all objects) to improve convergence. 
We train on 4 million samples from the prior, setting 10\% of the training data aside for validation, and stopping the training when the loss does not improve for 20 epochs. 

\subsubsection{Emission line fluxes}

Our model photometry has additional contributions from emission lines, so for any parameter vector $(\boldsymbol{\varphi}, z)$ we need to calculate them in addition to the base SPS predictions.
This extra computational load can again be alleviated with emulators.
However, in this case building emulators for individual emission lines and all bands would be prohibitive.
Instead, we adopt a different strategy.
We train an emulator to model the fluxes $\{ \ell_j \}$ on the rest-frame wavelength grid $\{ \lambda_j\}$, following the procedure of \cite{Alsing_2020}.
In our setting the size of the wavelength grid is the number of emission lines in \textsc{FSPS} (all $128$ lines in \citealt{Byler_2017}), so the accuracy of the emulator is easier to keep under control\footnote{For reference, a standard FSPS spectrum with the default MILES spectral library has $\sim 6\times 10^3$ wavelength points.}. 
To predict the photometric flux of an emission line, we need to implement \equref{eq:fluxdefinition}.
Because the band-passes $\{ W_b(\lambda) \}$ are in tabular form, it would be preferable to work with an interpolated or continuous approximation.
We find that fitting the sum of three generalized normal distributions to each $W_b$ offers a simple solution, since it is straightforward to fit to the tabulated data without any tuning. This approach delivers accurate and fast results for the final emission line fluxes.

\subsubsection{Emulator accuracy}

When comparing the initial and the emulator-reconstructed fluxes (computed from fluxes), we find that they are accurate at a level better than 1\% (99th percentile of the magnitude differences in the range spanned by the COSMOS2020 data below).
This will be comfortably covered by $\Sigma_b$, for example thanks to the $\gamma_b$ hyper parameter being greater than 0.01.
The accuracy of emulation of emission line fluxes is typically below $0.1\%$ (99th percentile) so the errors in this emulation can safely be ignored.

\begin{figure}
    \centering
    \hspace*{-4mm}\includegraphics[width=1.07\columnwidth]{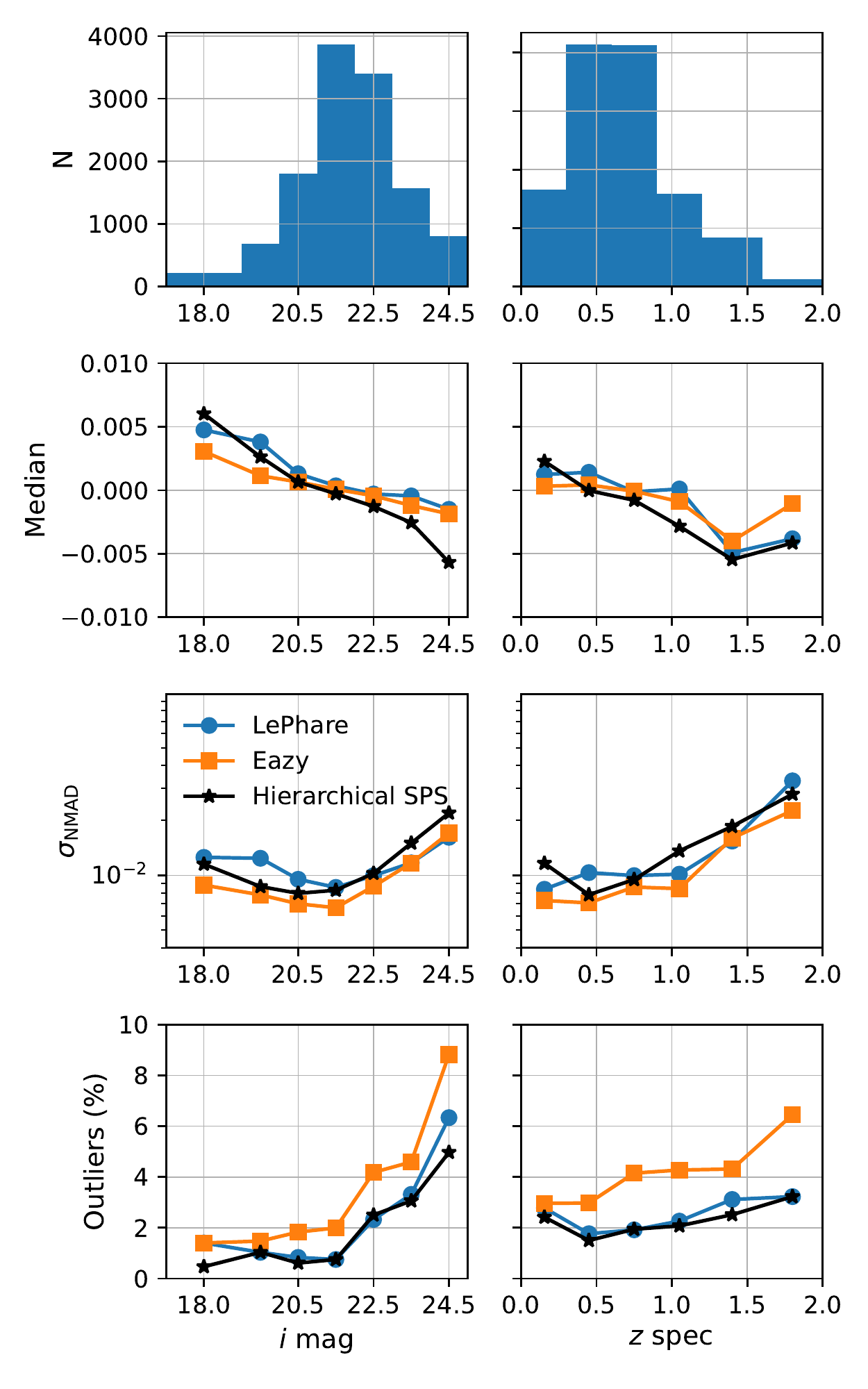}
    \caption{Photo-$z$ metrics calculated on our sample for our hierarchical model compared with the COSMOS2020 released {\Lephare} and {\eazy} \photoz's. We compute the median of $\Delta_z = ( z_\mathrm{spec} -  \hat{z}^\mathrm{MAP})/(1 + z_\mathrm{spec})$, as well as $\sigma_\mathrm{MAD}=1.48\times \mathrm{median}(|\Delta_z|)$, and the outlier fraction as the fraction of objects with $|\Delta_z|>0.15$.}
    \label{fig:photoz_metrics}
\end{figure}

\begin{figure*}
    \centering
    \hspace*{-0.4cm}\includegraphics[width=18.5cm]{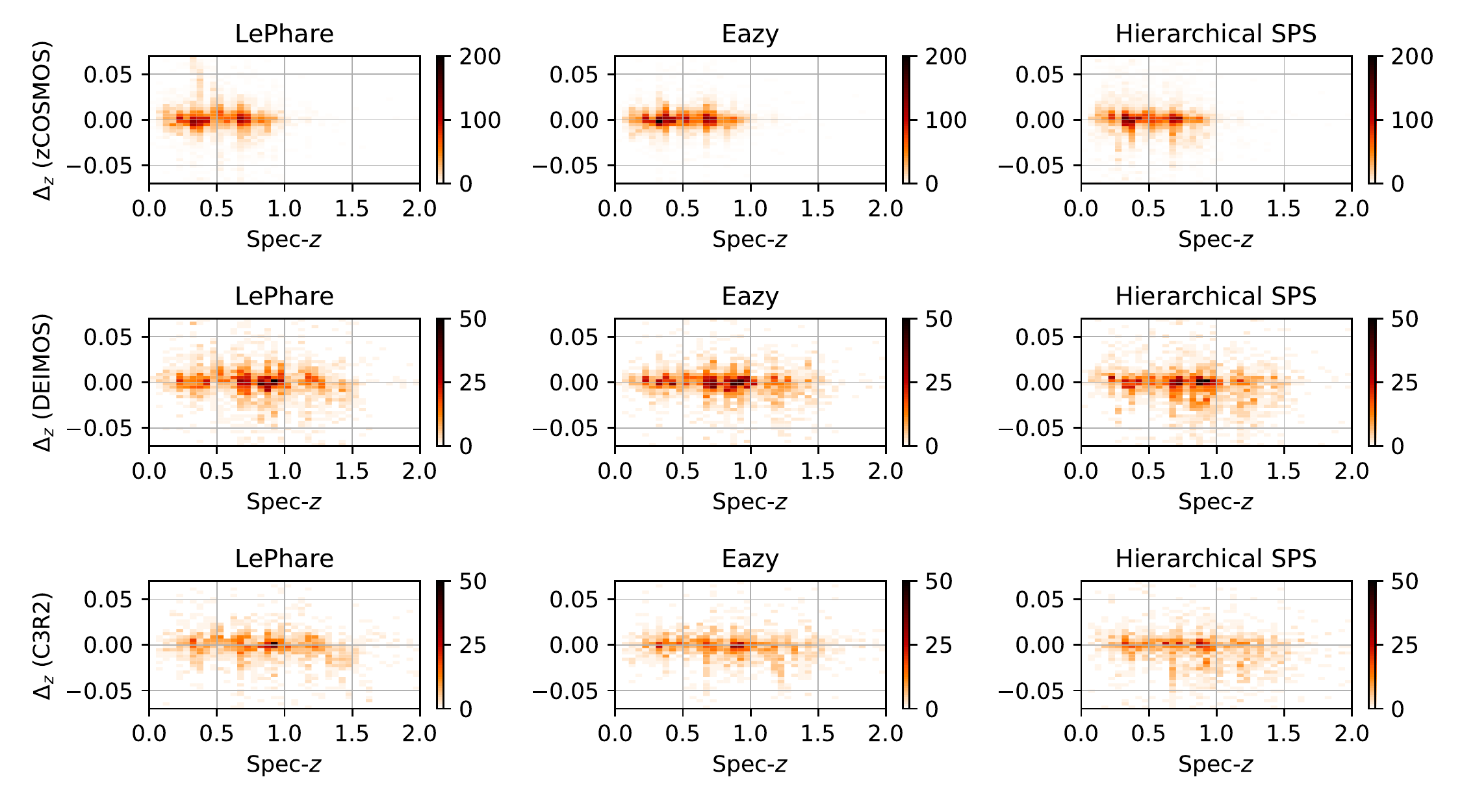}
    \caption{Differences between the maximum a-posteriori redshift estimates and the spectroscopic redshifts (divided by $1+z$) for the new model compared with the COSMOS2020 released {\Lephare} and {\eazy} redshifts. This highlights the areas where they achieve similar performance in terms of scatter and outliers. Splitting between the three sources of spectroscopic redshifts also reveals the differences in \photoz\  performance for samples with different selection, as described in \secref{sec:data}.
    }
    \label{fig:photoz_diffscatter}
\end{figure*}

\begin{figure*}
    \centering
    \hspace*{-0.4cm}\includegraphics[width=18.5cm]{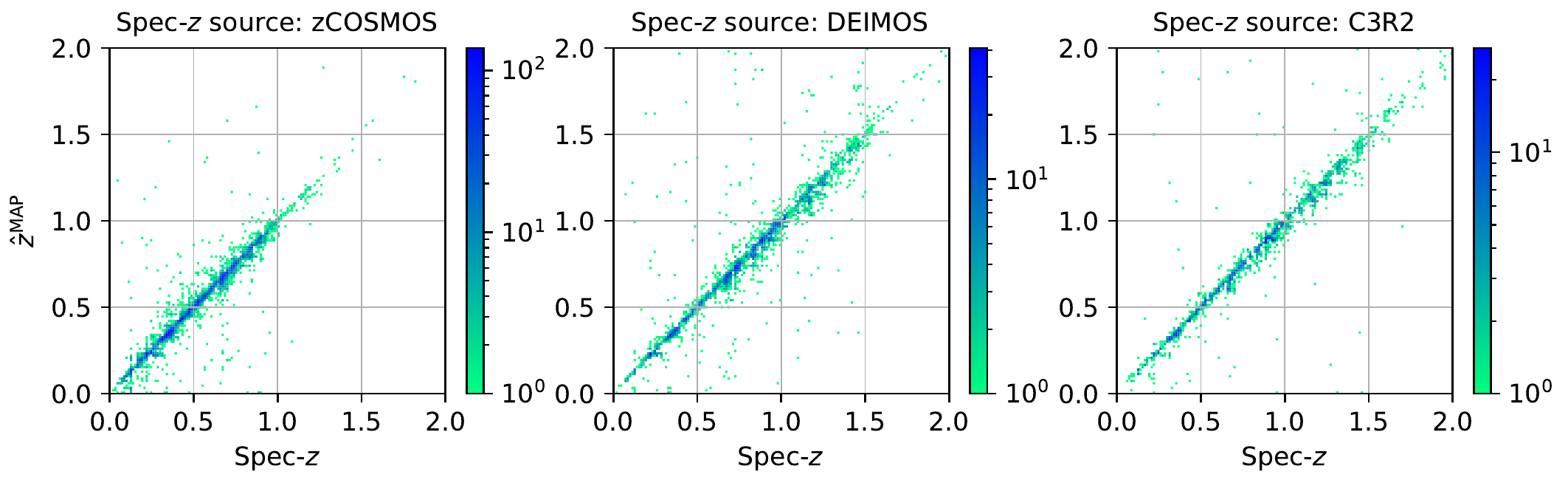}
    \caption{Maximum a-posteriori redshift estimates versus spectroscopic redshifts for the three spectroscopic sources comprising the data we analyze. Differences are best highlighted in \figref{fig:photoz_diffscatter}, whereas these scatter plots with a logarithmic color bar better highlight the overall distributions and the outliers (which are rare).
    }
    \label{fig:photoz_scatter}
\end{figure*}

\newpage
\section{Data} \label{sec:data}

We now construct a dataset that demonstrates that our framework is able to produce accurate photometric redshift inferences on a benchmark sample. This consists of selecting broad- and narrow-band photometry in the COSMOS field, and cross-matching with a selection of spectroscopic datasets to enable validation of inferred redshifts.

\subsection{Photometry}

\cite{Weaver_2022} presented COSMOS2020, the latest compilation of imaging in the COSMOS field.
Two sets of photometric measurements were performed on these data: \textit{Classic}, and \textit{Farmer}. 
For our case study we  choose \textit{Farmer}, which is based on the model-fitting software The Tractor\footnote{\url{https://github.com/dstndstn/tractor}}.
The photo-$z$'s published with this dataset were derived using {\eazy} \citep{Brammer_2008}, as well as {\Lephare} \citep{Ilbert_2006, Ilbert_2008} using slightly different subsets of bands.
While we will compare our results to both of these methods, we adopt the same subset of selected bands as {\eazy}; its SED templates are also derived using FSPS, thus providing a more direct comparison. This selection of bands exclude the Subaru Suprime-Cam broad-bands (shallower than other filters at similar wavelengths) and the GALEX bands (shallow and with broad PSFs) \citep{Weaver_2022}.
We apply the recommended `combined’ mask, which retains the deepest regions with the greatest number of available bands, and removes areas corrupted by bright stars and artifacts in all the relevant bands. 

This photometric dataset is prepared for further analysis using the code released with the COSMOS2020 data, which applies various flux corrections (including Galactic extinction) and unit conversions. 

\subsection{Spectroscopic redshifts}

We compile the positions and spectroscopic redshifts from three large campaigns covering the COSMOS field: 
\textbf{zCosmos-bright} \citep{Lilly_2007}, \textbf{DEIMOS} \citep{Hasinger_2018}], and \textbf{C3R2} \citep{Masters_2017, Masters_2019, Stanford_2021}.

{zCosmos-bright} \citep{Lilly_2007} is a highly representative sample of 20k bright ($i*\leq 22.5$) galaxies. 
{DEIMOS} \citep{Hasinger_2018} is a sample of $\sim 4 \times 10^3$ objects selected from a variety of input catalogs based on multi-wavelength observations in the field, and thus represents a diverse selection function.
Finally, the Complete Calibration of the Color-Redshift Relation (C3R2, \citealt{Masters_2017, Masters_2019, Stanford_2021}) is a multi-institution, multi-instrument survey targeting faint galaxies ($i\sim 24.5$) populating regions of color space under-sampled by other surveys, in order to provide validation data for upcoming weak lensing surveys.
We use data from the Data Releases 1, 2, and 3 ($\sim 4 \times 10^3$ objects).

For all three samples, we keep objects with quality flags 3 and 4.
We perform a spatial cross-match between the positions of these objects in the spectroscopic and photometric catalogs. 
Specifically, we keep the nearest unique photometric source within a radius of 0.6 arcseconds around each spectroscopic object, as recommended by \cite{Weaver_2022}. 
We also remove objects classified as stars by {\eazy} or {\Lephare}, as well as the few sources with spectroscopic redshift $>2.5$. 

We are left with $12,473$ objects: 6920, 3168, and 2043 for zCosmos-bright, DEIMOS, and C3R2, respectively.
This is the dataset on which we carry out our demonstration study.

This dataset combines three samples and so has a complicated selection function (\ie the probability, conditional on its intrinsic properties, that a galaxy appears in the sample).  
The selection function must be incorporated into any population level analysis, as is treated explicitly in the inference of redshift distributions presented by \cite{Alsing_2022}.  
But if the focus is on inference of object-level parameters (\ie individual galaxy redshifts) then selection effects can be ignored if the data on detected objects is sufficiently constraining: in the limit that the photometry determines the redshift (and other parameters) perfectly, the posterior is a delta function and hence completely decoupled from the population.  
While not strictly satisfied here - the high-quality photometry and broad wavelength coverage do yield finite parameter uncertainties - this is an excellent approximation.  We hence do not consider selection effects further.

\begin{figure}
    \centering
    \includegraphics[width=\columnwidth]{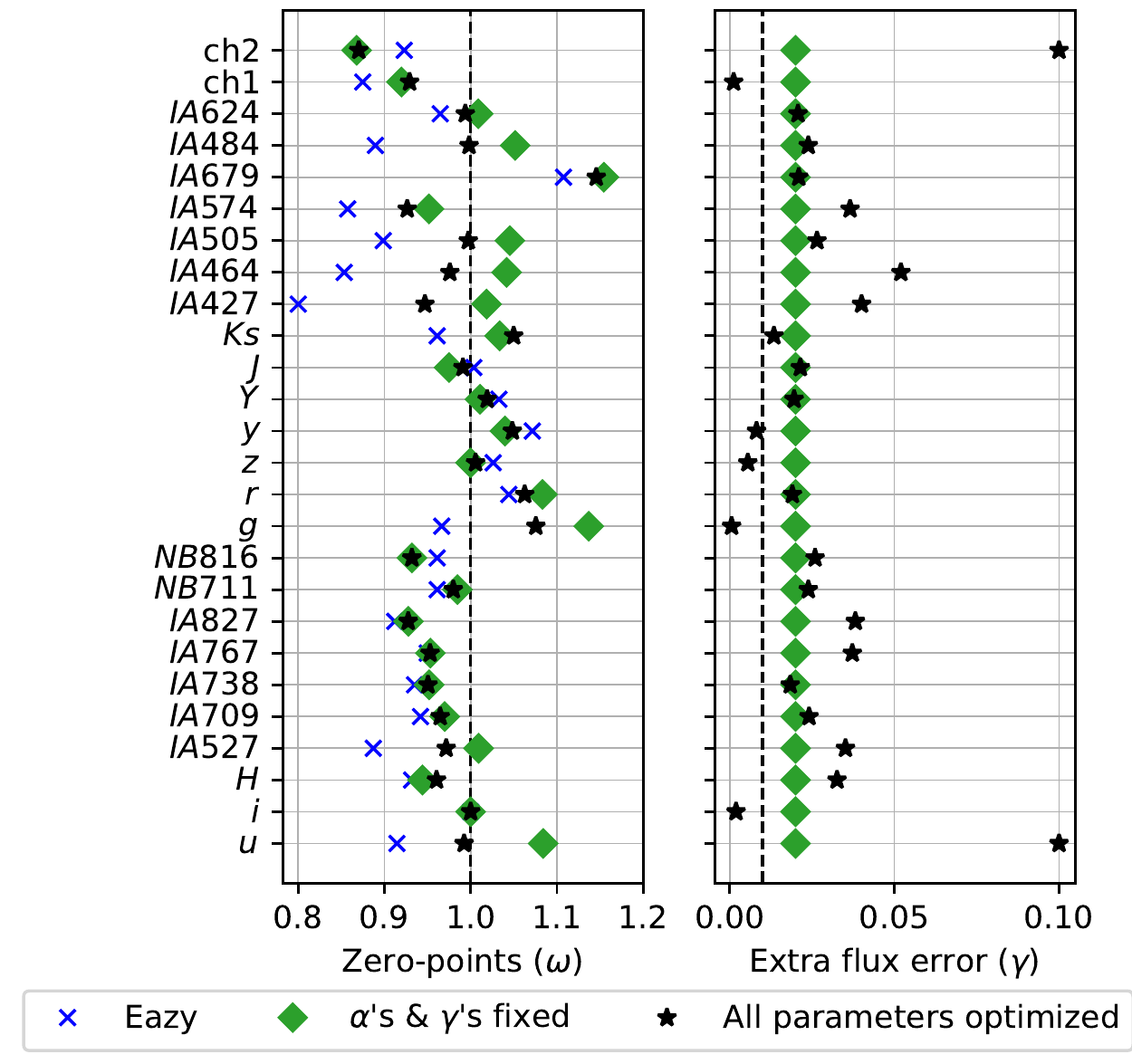}\\
    \includegraphics[width=\columnwidth]{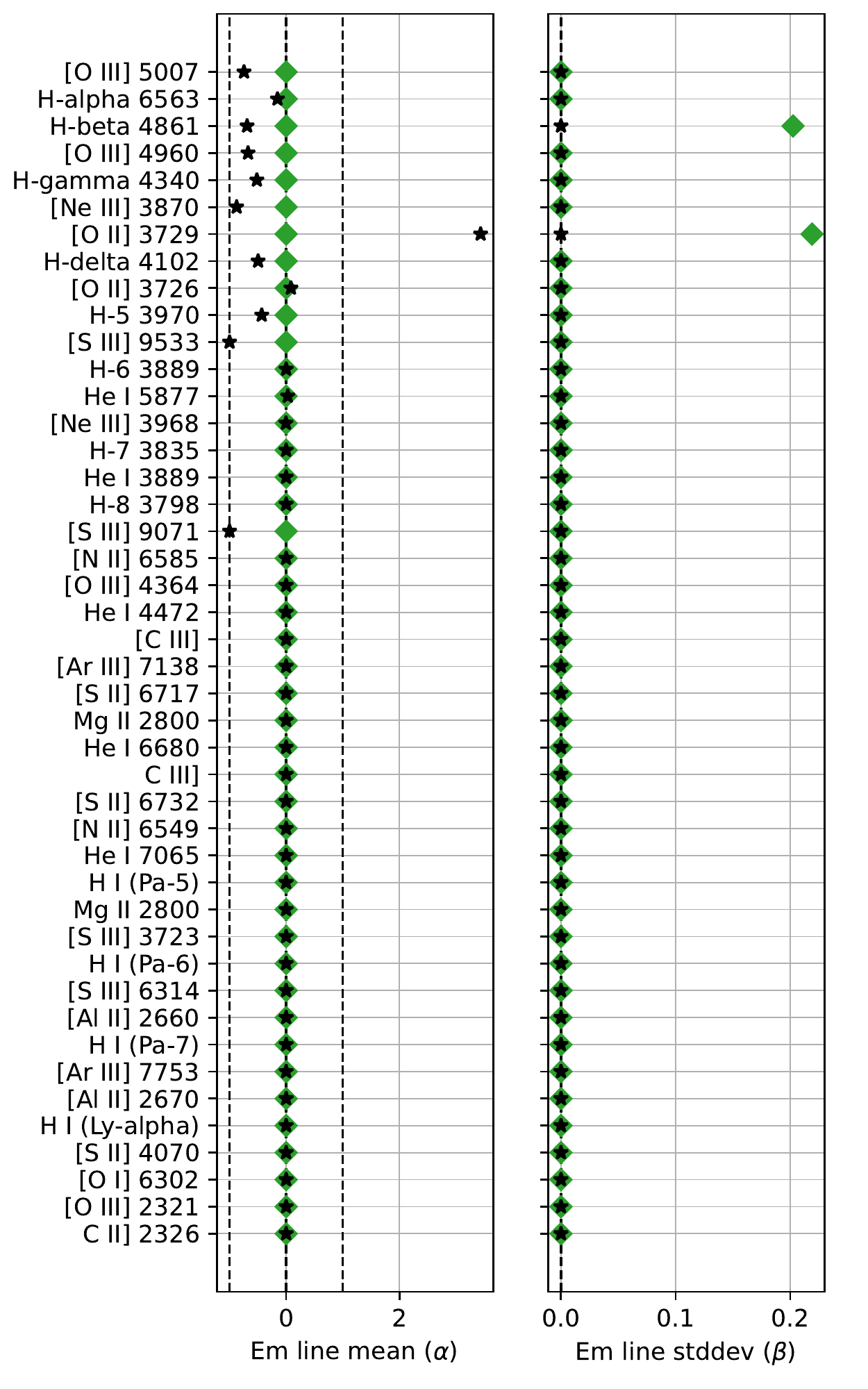}
    \caption{Hyper-parameters obtained with the optimization procedure of \secref{sec:optimization}. The black stars correspond to our best model, with all parameters optimized, while the green diamonds correspond to an additional model (used for robustness checks, see text for details) fixing the emission line offsets $\alpha_j=0\ \forall j$ and error floors $\gamma_b=0.01\ \forall b$.  
    }
    \label{fig:hyper_params}
\end{figure}

\section{Results} \label{sec:results}

We now evaluate the performance of our inference framework applied to the cross-matched dataset described in the previous section. In what follows, unless explicitly stated, the hyper-parameters of the population model ($\boldsymbol{\kappa}$) are not optimized, and are set to the fiducial values defined previously. 

In practice, to avoid degeneracies with other parameters of the model, we anchor the zero-point of the HSC $i$ band to unity, and define the others relative to this band.

\subsection{Emission lines}

We perform an initial fit to the data with our model with no contribution from lines, \ie setting $\alpha_j=\beta_j=0 \ \forall j$.
Thus, we optimize the zero-point offsets and uncertainty hyper-parameters $\{\omega_b, \gamma_b\}_{b=1, \cdots, \mathrm{N}_\mathrm{bands}}$, as well as the SPS parameters of each galaxy, with the procedure outlined in \secref{sec:optimization}. 
This consists of a run with redshifts fixed at the spectroscopic redshifts to estimate the hyper-parameters, and a second run now relaxing redshifts but fixing the hyper-parameters.
The resulting maximum a-posteriori redshift estimates,  $\hat{z}^\mathrm{MAP}$, are poor compared to the released COSMOS2020 \photoz's. 
We also observe sharp features in the redshift residuals $\hat{z}^\mathrm{MAP}-$spec-$z$ and in the flux residuals (data minus model at the best-fit parameters) binned in spec-$z$.
Many of these features correspond to strong emission lines, such as [O II], demonstrating that their luminosities are not correctly calibrated for our data set.
Re-calibrating them is straightforward within our framework.

We must first determine a suitable subset of emission lines (out of the $128$ implemented in FSPS by \citealt{Byler_2017}) to include in the model.
To achieve this we take the best-fit SEDs of our first run described above (at $z$=spec-$z$) and calculate the contribution of each emission line to the total photometric flux, \ie the flux ratio $L_{jb}/F_{b}$, for all objects, lines, and bands.
We then take the average of these values over objects and bands.
This procedure yields the lines which make the strongest contributions to the data at hand.
This set includes most of the lines with intrinsic large variance according to \cite{Byler_2017}.
Importantly, this procedure also filters out lines that are outside our wavelength coverage.
We are left with $44$ lines.
Note that nearby emission lines are not necessarily all well-resolved by the data, \ie each may only unequivocally contribute to a few objects.
This could be improved with more informative priors on the lines or on line-selection; however we do not expect this to have a significant impact on our redshift results.

In total, following this line-selection procedure, we have $2 \times \mathrm{N}_\mathrm{bands} + 2\times \mathrm{N}_\mathrm{lines}  = 140$ hyper-parameters (excluding $\boldsymbol{\kappa}$), and $15\times \mathrm{N}_\mathrm{objects} \sim 2 \times 10^5 $ `latent' SPS parameters. 
We now optimize all these parameters and extract maximum a-posteriori redshift estimates $\hat{z}^\mathrm{MAP}$. 

\subsection{Photo-$z$'s}\label{sec:photozs}

We evaluate the performance of this model using some commonly-used metrics.
For every object in our catalog, a spectroscopic redshift $z_\mathrm{spec}$ and a redshift estimate $\hat{z}^\mathrm{MAP}$ are available.
The main quantity of interest is the difference between the two, divided by the standard $1+z$ factor accounting for the expected scaling of the quality of the estimates due to the effect of redshift on wavelength,
\eqn{
    \Delta_z = \frac{ z_\mathrm{spec} -  \hat{z}^\mathrm{MAP} }{ 1 + z_\mathrm{spec}} \,.
}
We compute the mean, median, and standard deviation of $\Delta_z$ over our sample of galaxies.
For the standard deviation, an estimator more robust to outliers than the sample variance is $\Sigma_\mathrm{NMAD}$, the median absolute deviation (the median of $|\Delta_z|$) multiplied by $1.48$.
Finally, the outlier fraction is defined as the percentage of objects  with $|\Delta_z|>0.15$.

\figref{fig:photoz_metrics} shows these metrics, calculated in redshift and magnitude bins (in the reference HSC $i$ band, which is not affected by zero-points), for our framework as well as the publicly-released COSMOS2020 {\Lephare} and {\eazy} redshifts.

Overall, our approach yields comparable levels of bias (as measured by the median of $\Delta_z$), fewer outliers, and a slightly larger level of scatter (as measured by $\sigma_\mathrm{NMAD})$. 
The comparison with {\Lephare} is not straightforward since it uses a slightly different set of bands.
On the other hand, we use the same set of bands as {\eazy}, also based on templates derived from a grid of parameters using a SPS model. 
{\eazy} includes a magnitude-redshift prior as well as a template error function \citep{Brammer_2008}.
These differences may be responsible for the small improvement in  $\sigma_\mathrm{NMAD}$. Note that we have not attempted to tune or improve our model in order to carry out redshift inferences on these data, beyond the procedure described above. 


While these {\photoz} metrics are informative summaries of the results, it is also interesting to examine the distributions of redshift estimates themselves. 
The residuals $\Delta_z$ are shown in \figref{fig:photoz_diffscatter}, while \figref{fig:photoz_scatter} also shows a conventional comparison of the estimates and the spectroscopic redshifts.
These confirm the results captured by the metrics above.
This also highlights that the redshift estimates are noisier for the DEIMOS and C3R2 samples because these surveys targeted populations typically under-represented in other spectroscopic campaigns (fainter, bluer, higher redshift). These populations also present greater challenges in terms of accurate \photoz~estimation.

\begin{figure*}
    \centering
    \includegraphics[width=18cm]{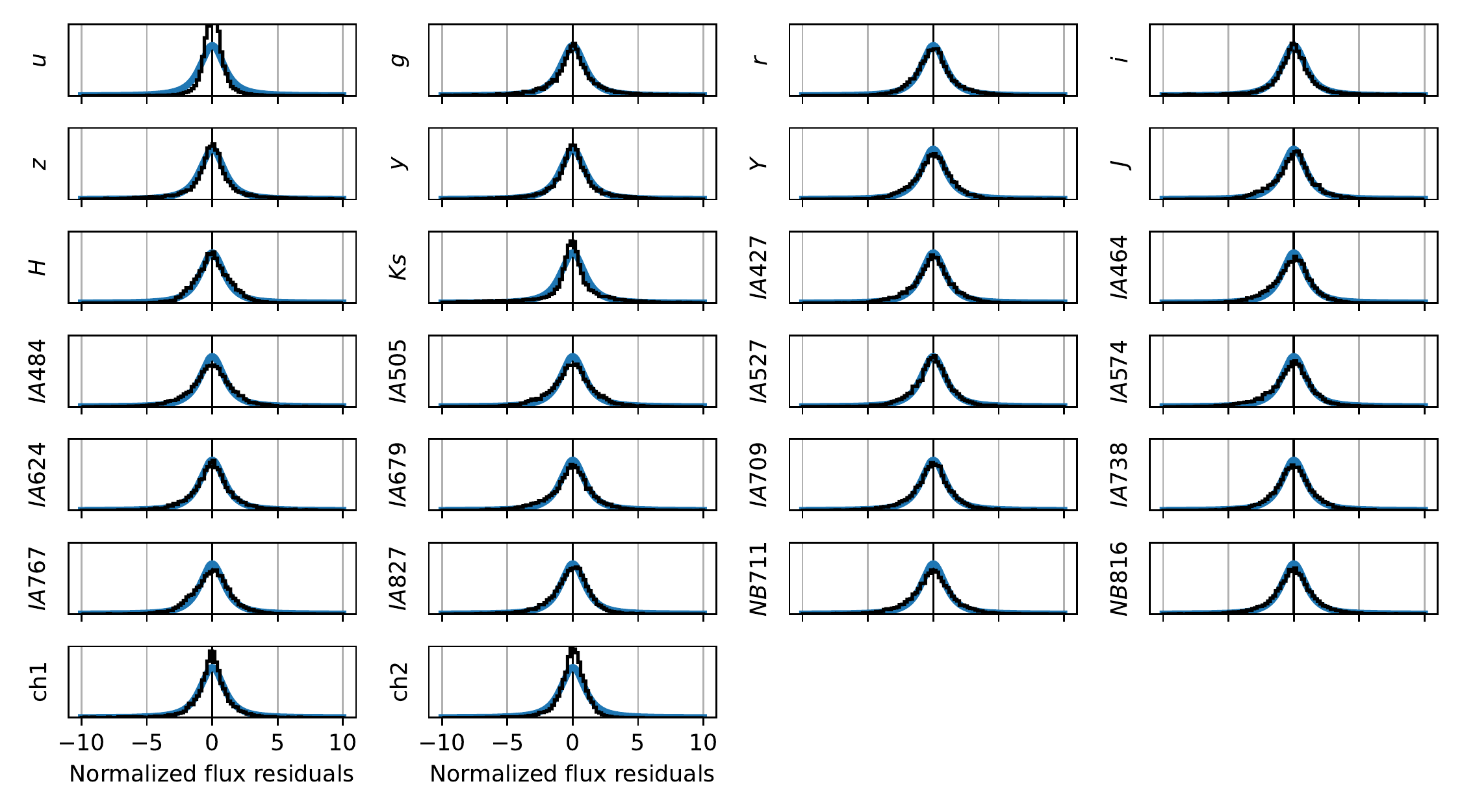}
    \caption{Flux residuals (black), as defined in \equref{eq:residuals}, compared with the likelihood function of \equref{eq:likelihood} in blue, showing a satisfactory agreement between the data and the model predictions.}
    \label{fig:residuals}
\end{figure*}

\subsection{Hyper-parameters}

In this section, we investigate the results of our optimization procedure for the model's hyper-parameters, and their implications for the robustness of the data-model. 
The hyper-parameters resulting from the optimization are shown in \figref{fig:hyper_params}.  
We also show the result of a run with the noise and emission line offsets set to fiducial values $\gamma_b=0.01\ \forall b$ , $\alpha_i=0 \ \forall i$.
Throughout this section we will also comment on robustness checks obtained by running variations of these models.

\subsubsection{Zero-point offsets}

The inferred zero-points $\{\omega_b\}$ are at the level of a few percent in most bands.
They are well-constrained by the data and robust to changes in the model.
The biggest changes are shifts in the intermediate bands caused by emission line offsets $\alpha_i$, as shown in \figref{fig:hyper_params}. 
These results are qualitatively consistent with the \eazy\ zero-points released with COSMOS2020. 
This is not a surprise given that we use the same set of bands and that \eazy~also involves a grid of SPS templates. 
Some possible explanations for the disparities are differences in the SPS modeling, the approximation resulting from using a grid as opposed to a continuous model, or even the fact that {\eazy} calibrates the zero-points based on the whole COSMOS2020 dataset (not only objects with spectroscopic redshifts). 

\subsubsection{Examination of photometric uncertainties}

\figref{fig:hyper_params} shows that the inferred values of flux uncertainties $\gamma_b$ are at the level of a few percent in most bands: smaller for the broad bands (which are typically deeper), and larger for intermediate and narrow bands (noisier and also more sensitive to the details of the SED modeling, such as emission lines). 
The $u$ and ch$2$ bands are the most noisy, consistent with previous findings \citep{Laigle_2016, Weaver_2022}.
Just as for zero-points, the inferred values are mostly stable when switching other hyper-parameters on and off. 
However, there is no uncertainty contributed by emission lines in this model (bottom right panel), indicating that the band-to-band uncertainty is preferred by the data. If we set the band-to-band uncertainty to 1\% (green diamonds), then emission lines contribute to the uncertainty budget (via non-zero $\beta$) to compensate, as expected.
Thus, we conclude that the noise model performs well, but that emission line uncertainty is simply not required in our full model for this dataset.

\subsubsection{Emission line offsets}

We find that including extra contributions from emission lines significantly improves the quality of the derived {\photoz}’s.
This is consistent with previous findings \cite{Ilbert_2006, Ilbert_2008, Brammer_2008, Alarcon_2021}.
Note that in \figref{fig:hyper_params} emission lines are ordered by their contribution relative to the total broadband flux, following the procedure described above.
Offsets are primarily given to the strongest lines, with some of them being almost entirely cancelled ($\alpha \approx -1$).
The Laplace prior is effective at keeping most line offsets at zero, unless they bring a significant improvement to the fits. 
We also note that the interpretation of the hyper-parameters is conditioned on the sample at hand: changing the sub-set of COSMOS2020 galaxies analysed may change the resulting values of zero-points and emission line offsets, for example.

\subsection{Population model}

We perform one additional run also optimizing the hyper-parameters of the population model, $\boldsymbol{\kappa}$. 
We find that the data do not significantly perturb the fiducial values originally adopted; nor do these parameters appear degenerate with other parameters of the model. This demonstrates the robustness of the model to the prior, and additionally shows that varying these hyper-parameters does not lead to significantly improved fits.
However, it should be noted that we have not included selection effects in our model. If we had found that the inferred values had significantly moved from the fiducial values, we would have been unable to determine whether this had been caused by incomplete modeling of the selection process \citep{Alsing_2022}. However in this setting, as we have previously argued, the selection effects should be negligible so this concern does not arise.

\subsection{Residuals}

Finally, we check the ability of the presented model to explain the statistical properties of the data. We examine the distribution of the flux residuals, defined as 
\eqn{
    r_b = \frac{\hat{F}_b -  \omega_b F_b}{
\sqrt{\sigma_b^2 + (\omega_b\Sigma_b)^2}} \, . \label{eq:residuals}
}
These are shown in \figref{fig:residuals}.
The blue lines show our likelihood function, the Student's-t distribution described in \equref{eq:likelihood}.
There is good agreement between the chosen likelihood and the residuals. We have checked that this agreement significantly deteriorates when all the hyper-parameters are not optimized (not shown here).
The deviations in the $u$ and ch2 bands are likely to be due to residual data systematics which are not well-described by our model.

When repeating this analysis with a normal likelihood function, we find that the residuals are worse and provide a poor fit of the likelihood function, despite optimizing the other components of the model. 
This setting also results in worse redshift estimates.
This demonstrates the presence of outliers, and shows that accommodating them with a suitable likelihood function can significantly improve the ability of a model to fit complicated data.


\section{Discussion} \label{sec:discussion}

Our results  demonstrate that the  hierarchical SPS inference framework presented here (and in \citealt{Alsing_2022}) delivers accurate redshifts on a benchmark dataset, therefore passing a stringent validation test.
We now discuss possible extensions which relax some of its assumptions.

The model itself could be made more sophisticated in various ways in order to better suit specific applications and increase the accuracy of the \photoz's further.
Such extensions include a stellar model, which could straightforwardly be incorporated to allow star-galaxy separation, akin to what is done in \eazy\ and \Lephare\ with stellar templates. 

In the future it will be critical to include effects which are currently neglected but affect the data (and in turn, photometric redshifts) at significant levels for LSST, such as airmass-dependent fluxes (\citealt{Graham_2018}).
Furthermore, extending our model to include spatial information \citep{S_nchez_2018, Jasche_2019, Alarcon_2020, S_nchez_2020} could provide a powerful probe of the connection between galaxies and dark matter, and exploit more of the information offered by photometric surveys. 

Our chosen optimization strategy, paired with the posterior distribution taken as a loss function, minimizes the residuals between the measured and the model fluxes.
Other choices may be possible, especially when including spectroscopic redshifts.
For example, one could explicitly minimize the residuals between photometric (inferred) and spectroscpic redshifts, although this will render the results more sensitive to selection effects in the data.

We neglected the uncertainties on hyper-parameters and SPS parameters.
This is because the extensive wavelength coverage of the \bl{26-band COSMOS2020 data, the small flux uncertainties, and the availability of spectroscopic redshifts combine to yield well-constrained galaxy SEDs and in turn hyper-parameters.
Nevertheless, there are fundamental degeneracies expected between some of the SPS parameters \citep{Leja_2017, Leja_2019}, as well as with the redshift when it is not fixed to a spectroscopic value.
Selection effects may also be important, and can be accounted for in the posterior distributions as described in \cite{Alsing_2022}.
In analyzing datasets where the uncertainties on hyper-parameters or SPS parameters (including redshift) are not negligible (\eg broadband-only surveys like the LSST), one can resort to the variety of Bayesian inference techniques (\ie MCMC, variational inference, Laplace approximation, simulation-based inference) available for hierarchical models. 
Since the impact of selection cuts and parameter uncertainties is strongly dependent on the specific data and models being considered, we defer further discussion and the development of effective inference strategies to future work.} 



\section{Conclusion} \label{sec:conclusion}

We have presented a hierarchical model to infer redshift and other intrinsic galaxy properties from photometric data.
This approach makes it possible to self-consistently encode knowledge of (1) galaxy formation and evolution via a population model, (2) astrophysics via SEDs predicted with stellar population synthesis, (3) observational effects via a noise model and a likelihood function.
By formulating the inference within a Bayesian hierarchical model, we are able to parameterize additional sources of bias and uncertainty in each of these components, and to solve for them self-consistently.
Thus, any manual tuning or inversion methods can be avoided, with the added benefit that it is possible to set informative priors on these additional terms.
 
Our {\photoz}'s (as measured with traditional metrics on redshift point estimates) are competitive with the publicly released {\photoz}'s from COSMOS2020 \citep{Weaver_2022}. 
While this is a benchmark test on relatively bright, low-redshift sample of spectroscopically-confirmed galaxies, these results demonstrate the power and flexibility of our methodology, and motivate its further development towards addressing the challenging setting presented by stage IV cosmological surveys. 
We made our inference tractable using neural emulators to accelerate model calls \citep{Alsing_2020}.

In future applications of our framework, we will relax some of the modeling assumptions made here.
In particular, we demonstrate how to incorporate selection in the companion paper \citep {Alsing_2022}. 
Comparing inferences from multiple SPS and population models will also provide valuable consistency checks.
In fact, refining the SED and population models directly from the upcoming survey data is a promising avenue for directly probing the underlying (astro)physics using large amounts of untapped statistical power.
Leveraging hybrid data sets and incorporating additional data available (\eg spec-$z$'s, or even full spectra) will be instrumental for correctly separating the different physical effects at play.
This is straightforwardly included in hierarchical models (see \eg \citealt{Alarcon_2021, Nagaraj_2022}). 
Together, these attributes make the methodology presented here well-suited for addressing the challenges of upcoming photometric surveys and the robust scientific exploitation of their data.
\\
\\
\\
\textbf{Author contributions.} 
{\bf BL:} Conceptualization, methodology, software, validation, formal analysis, writing - original draft.
{\bf JA:} Conceptualization, methodology, software, validation, writing - review \& editing.
{\bf HVP:} Conceptualization, methodology, validation, writing - review \& editing, funding acquisition.
{\bf DM:} Conceptualization, methodology, validation, writing - review \& editing, funding acquisition.
{\bf JL:} Conceptualization, methodology, validation, writing - review \& editing.

\textbf{Acknowledgements.} 
We thank George Efstathiou for valuable input during the course of this project, and John R.\ Weaver for assistance with the COSMOS2020 data.
We also thank Konrad Kuijken, Hendrik Hildebrandt, Angus Wright, and Will Hartley for useful discussions.
BL is supported by the Royal Society through a University Research Fellowship. This project has received funding from the European Research Council (ERC) under the European Union’s Horizon 2020 research and innovation programme (grant agreement no. 101018897 CosmicExplorer). This work has also been enabled by support from the research project grant ‘Understanding the Dynamic Universe’ funded by the Knut and Alice Wallenberg Foundation under Dnr KAW 2018.0067. JA, HVP and DJM were partially supported by the research project grant “Fundamental Physics from Cosmological Surveys” funded by the Swedish Research Council (VR) under Dnr 2017-04212. The work of HVP was additionally supported by the Göran Gustafsson Foundation for Research in Natural Sciences and Medicine. HVP and DJM acknowledge the hospitality of the Aspen Center for Physics, which is supported by National Science Foundation grant PHY-1607611. The participation of HVP and DJM at the Aspen Center for Physics was supported by the Simons Foundation.




\bibliography{bib}{}
\bibliographystyle{aasjournal}



\end{document}